\documentclass[prb,preprint,showpacs]{revtex4}
\usepackage[dvips]{graphicx}
\usepackage{color}
\usepackage{bm}
\usepackage{epsfig}
\usepackage{amsmath,amssymb,amscd}

\def\beq{\begin{equation}}
\def\eeq{\end{equation}}
\newcommand{\mbold}[1]{\mbox{\boldmath $ #1 $}}
\begin{document}
\title{
Nagaoka ferromagnetism in large-spin systems\\
        -Fermion and Boson systems-- 
}
\author{Seiji Miyashita},
\author{Masao Ogata}
\affiliation{
Department of Physics, Graduate School of Science,
University of Tokyo, Bunkyo-ku, Tokyo 113-0033, Japan
}
\affiliation{
CREST, JST, 4-1-8 Honcho Kawaguchi, Saitama, 332-0012, Japan}
\author{Hans De Raedt}
\affiliation{Department of Applied Physics,
Zernike Institute of Advanced Materials,
University of Groningen, Nijenborgh 4, NL-9747 AG Groningen, The Netherlands}

\begin{abstract} 
We study magnetic properties of itinerant quantum magnetic particles described by
a generalized Hubbard model with large spin ($S>1/2$) which may be realized in
optical lattices of laser-cooled atom systems. 
In fermion systems (half-integer spins), 
an extended form of Nagaoka ferromagnetism may be realized. 
However, as novel aspects of the large spin cases, we found that the condition 
on the lattice connectivity is more stringent than in the case of $S=1/2$ particles, 
and that the system shows a peculiar degenerate structure of the ground state
in which the ferromagnetic state is included. 
In contrast, it turns out that the ground state of
itinerant bosonic systems (integer spins) has
a degenerate structure similar to that of fermion system with $S>1/2$
regardless of the shape, connectivity or filling of the lattice, and 
that the state with the maximum total spin is always one of the ground states.
\end{abstract}



\maketitle


\section{Introduction}
\hspace*{\parindent}

The origin of magnetism is attributed to the quantum mechanical interaction 
of particle swhich carry spin.
In so-called localized spin systems, the Pauli principle plays an essential role
and the magnetic interaction is expressed as the Heisenberg 
interaction~\cite{Heisenberg}, 
the exchange integral between atoms being the dominant term.
Not only the two-spin interaction but also multi-spin interactions
may contribute to give rise to exotic magnetic states. 
In particular, various effects of the multi-spin interaction have been reported 
in $^3$He~\cite{multispin}.
For itinerant electron systems, the origin of the magnetic order, 
and of ferromagnetic order in particular, has been studied extensively. 
Itinerant electron systems are often described by tight binding models 
such as the Hubbard model~\cite{Hubbard}.
Nagaoka pointed out that the ground state of the Hubbard model
may be a ferromagnetic state (Nagaoka ferromagnetic state)
if the number of electrons is reduced by one from 
half-filling~\cite{nagaoka,Putikka},
where half-filling means that the number of particles is the same 
as the number of the lattice sites.
The Nagaoka ferromagnetic state is established if the so-called 
``connectivity condition''
on the lattice is satisfied~\cite{nagaoka,Watanabe1}. 
In the Nagaoka ferromagnetic state, the energy of the system is minimized 
when the total spin of the system takes the maximum value. 
On the other hand, the ground state of the half filled system
is a singlet state, that is its total spin is zero. 
Elsewhere, we have studied how the magnetic state changes between these two states 
when an electron is removed from the system and
demonstrated an adiabatic change between these states~\cite{adiabatic}. 

Magnetism has primarily been studied in electron systems in which the spin $S=1/2$.
However, recently developments in the field of laser-cooled atomic systems
have opened new possibilities to realize artificial tight-binding quantum system 
such as the Hubbard model~\cite{optical-lattice,takahashi}.
In contrast to electron systems, 
in optical lattices the spin of trapped atoms is not necessarily $S=1/2$ but
can take larger half-integer or integer values. 
In the latter case, the system contains bosons, not fermions,
and it is therefore of interest to study itinerant magnetism 
of systems with $S>1/2$. 
The present paper presents the results of such systems.

\section{Model}

We consider a tight binding model of the form
\beq
{\cal H}=-t\sum_{<ij>,M}\left(c_{i,M}^{\dagger}c_{j,M}^{\phantom{\dagger}}
+c_{j,M}^{\dagger}c_{i,M}^{\phantom{\dagger}}\right)
+\sum_{i=1}^NU(n_{i,M}),
\label{model}
\eeq
where $c_{i,M}^{\dagger}$ and $c_{i,M}^{\phantom{\dagger}}$ are 
the annihilation and creation operators of 
a particle (fermion or boson depending on its spin) of the magnetization $M$ 
at a site $i$, respectively, $n_{i,M}$ is the number operator
\beq
n_{i,M}=c_{i,M}^{\dagger}c_{i,M}^{\phantom{\dagger}},
\eeq 
for the particle of spin $S$, $M=-S,-S+1,\ldots,S$,
and $U(n_{i,M})$ represents the on-site repulsive interaction. 
For the system with $S=1/2$, we take for $U(n_{i,M})$ the standard form
\beq
U(n_{i,M})=U_0n_{i,-1/2}n_{i,1/2},
\eeq
and for larger $S>1/2$ we extend it as
\beq
U(n_{i,M})=\frac{1}{2}U_0 N_i(N_i-1),  
\eeq
where $N_i=\sum_M n_{i,M}$ and $U_0$ is assumed to be positive.
Obviously, the interaction $U(n_{i,M})$ increases the energy whenever
a site is occupied by more than one particle.

In order to study magnetic properties of the system, 
we introduce spin operators for the magnetization
$\mbold{S}=(S_i^x,S_i^y,S_i^z)$ of the particles, where
\beq
S_i^z=\sum_{M=-S}^S Mn_{i,M}.
\eeq
Hereafter, we use the operators 
$S_i^+=S_i^x+iS_i^y$ and $S_i^-=S_i^x-iS_i^y$
which are expressed in terms of $c_{i,M}^{\dagger}$ and $c_{i,M}^{\phantom{\dagger}}$: 
\beq
S^+_i=\sum_M\sqrt{(S-M)(S+M+1)}c_{i,M+1}^{\dagger}c_{i,M}^{\phantom{\dagger}}
\label{sp}
\eeq
and
\beq
S^-_i=\sum_M\sqrt{(S+M)(S-M+1)}c_{i,M-1}^{\dagger}c_{i,M}^{\phantom{\dagger}},
\label{sm}
\eeq
where $S$ is the total spin of each particle, 
and $M$ is the $z$-component of the magnetization.
To discuss the magnetic properties of the states,
we adopt the usual notation $|S,M\rangle$ where  
$S$ is the total spin $S$ and $M$ is the magnetization.
The action of the operators Eqs.~(\ref{sp}) and (\ref{sm}) 
on the state $|S,M\rangle$ is given by the relations 
\beq
S_i^+|S,M\rangle=\sqrt{(S-M)(S+M+1)}|S,M+1\rangle
\eeq
and
\beq
S_i^-|S,M\rangle=\sqrt{(S+M)(S-M+1)}|S,M-1\rangle.
\eeq

In order to explicitly compute matrix elements of the operators, 
it is convenient to introduce orthornormal basis states.
In the case of fermion systems, we adopt the form
\beq
|\Psi\rangle=c_{i,M}^{\dagger}\cdots c_{j,M'}^{\dagger}|0\rangle,
\label{Fstate}
\eeq
where $i\ge j$, and $M>M'$ if $i=j$. 
With this definition, the operations Eqs.~(\ref{sp}) and
(\ref{sm}) do not change the order of creation operators in Eq.~(\ref{Fstate}).
In the case of bosons, more than two particles with the same $M$ can occupy the same site
and the normalized basis states take the form
\beq
|\Psi\rangle=
\frac{                                                
(c_{i,M}^{\dagger})^{n_{i,M}}\cdots (c_{j,M'}^{\dagger})^{n_{j,M'}}|0\rangle
}{
\sqrt{n_{i,M}!}\cdots\sqrt{n_{j,M'}!}
}
\equiv |n_{i,M}\rangle\cdots |n_{j,M'}\rangle.
\label{Bstate}
\eeq
For boson systems, the order of $i$ and $j$ and $M$ and $M'$ is not relevant.

The total spin of the whole system is given by
\beq
\mbold{S}^2=(S^x)^2+(S^y)^2+(S^z)^2=\frac{S^+S^-+S^-S^+}{2}+(S^z)^2, 
\eeq
where
$
S^x=\sum_iS_i^x, \quad S^x=\sum_iS_i^x, \quad {\rm and} \quad S^x=\sum_iS_i^z,
$
and $S^{\pm}=\sum_i{S_i^x \pm iS_i^y}$. 
We denote the value of the total spin of the system by $S_{\rm tot}$, i.e., 
$S_{\rm tot}(S_{\rm tot}+1)=\langle\mbold{S}^2\rangle$.

\section{Conservation of number of particles of different spins and ground-state
degeneracy}

It should be noted that the Hubbard Hamiltonian conserves 
number of particles
\beq
n_M=\sum_{i=1}^N n_{i,M}
\eeq
for each $M$.
To describe the set of $n_M$ of states, 
it is convenient to introduce the notation
\beq
\{n_M \}= ( n_S,n_{S-1},\cdots,n_{-S} ),
\eeq
where $\sum_{M=-S}^{+S} n_M=N$.
It is important to note that except for $S=1/2$,
the operator $S_i^+S_j^-+S_i^-S_j^+$ changes the set $\{n_M \}$.
For example, if $S=1$ application
of $S_i^+S_j^-+S_i^-S_j^+$ to states in the set $(n_1=0,n_0=2,n_{-1}=0)$,
created states in the set $(n_1=1,n_0=0,n_{-1}=1)$.
Thus, except for $S=1/2$, the 
matrix element  $\langle \{ n'_M \} |S_i^+S_j^-+S_i^-S_j^+| \{ n_M \}\rangle$
can be nonzero even if $\{n_M \}\not= \{ n'_M \}$.

Thus, even though the Hamiltonian ${\cal H}$ has the SU(2) symmetry, 
conserves the total spin, 
and therefore commutes with $S^{\pm}$, 
the operation of $S_i^+S_j^-+S_i^-S_j^+$ 
on each state with given $\{n_M \}$ must be treated carefully.

Let us denote by $|G,M\rangle$ the ground state in the space 
with the magnetization $M$.
Because the Hamiltonian ${\cal H}$ and $S^{\pm}$ 
commute with each other, $S^{-}|G,M\rangle$ is also a ground state because
\beq
{\cal H}S^{-}|G,M\rangle=S^{-}{\cal H}|G,M\rangle=E_GS^{-}|G,M\rangle.
\eeq
However, the set of numbers $\{n_M \}$ is not necessarily
conserved, that is
$$
S^-{\cal H}|G,M,\{n_M \}\rangle=E_GS^-|G,M,\{n_M \}\rangle$$
\beq
=E_G\sum_{\{n'_M \}}a_{\{n'_M \}}|G,M-1,\{n'_M \}\rangle,
\eeq
where $|G,M,\{n_M \}\rangle$ denotes one of 
basis states with fixed $\{n_M \}$ that contribute
to the expansion of $|G,M\rangle$ in terms of basis states.
The energy of this state is given by
$$
E_G=
\frac{  
\langle G,M,\{n_M \}|S^+ {\cal H}S^- |G,M,\{n_M \}\rangle
}{
\langle G,M,\{n_M \}|S^+ S^- |G,M,\{n_M \}\rangle
}
$$ 
\beq
=\frac{  
\sum_{\{n'_M \}}|a_{\{ n'_M \}}|^2
\langle G,M-1,\{n'_M \}|{\cal H} |G,M-1,\{n'_M \}\rangle}{
\sum_{\{n'_M \}}|a_{\{n'_M \}}|^2\langle G,M-1,\{n'_M \}|G,M-1,\{n'_M \}\rangle
}.
\label{Eg}
\eeq
Because 
\beq
\frac{
\langle G,M-1,\{n'_M \}|{\cal H} |G,M-1,\{n'_M \}\rangle
}{
\langle G,M-1,\{n'_M \}|G,M-1,\{n'_M \}\rangle}
\ge E_G,
\eeq
in order to satisfy the relation (\ref{Eg}), the state in each set of
numbers $\{n'_M \}$ must be the ground state, i.e.
\beq
\frac{ 
\langle G,M-1,\{n'_M \}|{\cal H} |G,M-1,\{n'_M \}\rangle
}{
\langle G,M-1,\{n'_M \}|G,M-1,\{n'_M \}\rangle}= E_G,
\eeq
and thus
\beq
{\cal H} |G,M-1,\{n'_M \}\rangle=E_G|G,M-1,\{n'_M \}\rangle.
\label{HEG}
\eeq
This shows that the same ground state energy
is found for all sets $\{n'_M \}$
except if $a_{\{n'_M \}}=0$.

\section{Ground state of a fermion system with $S=\frac{3}{2}$}

\subsection{Subspaces due to conservation of particle number of each $M$ }

\begin{figure}[t]
$$
\begin{array}{cc}
 \includegraphics[width=.20\textwidth]{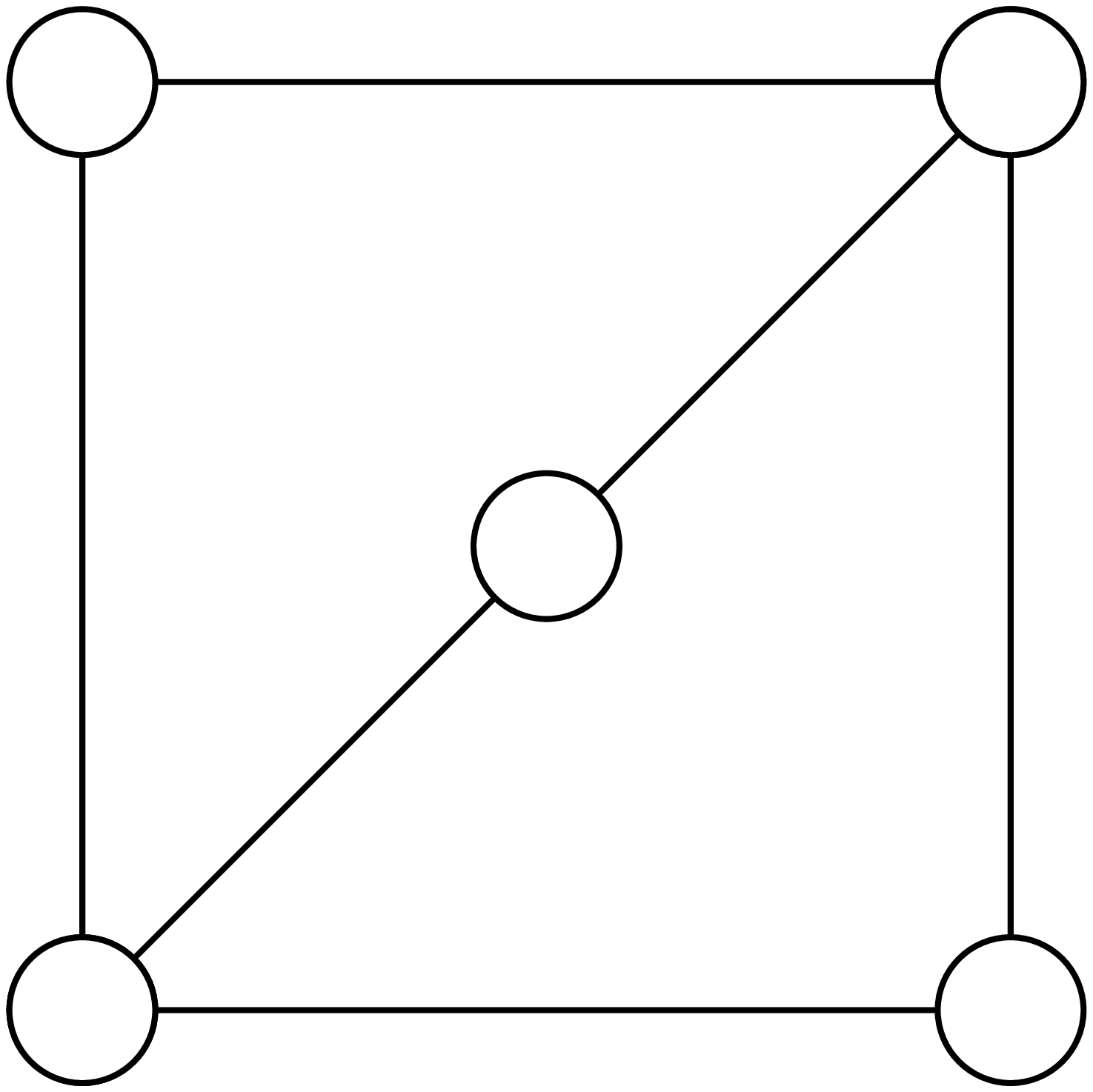}&
 \includegraphics[width=.20\textwidth]{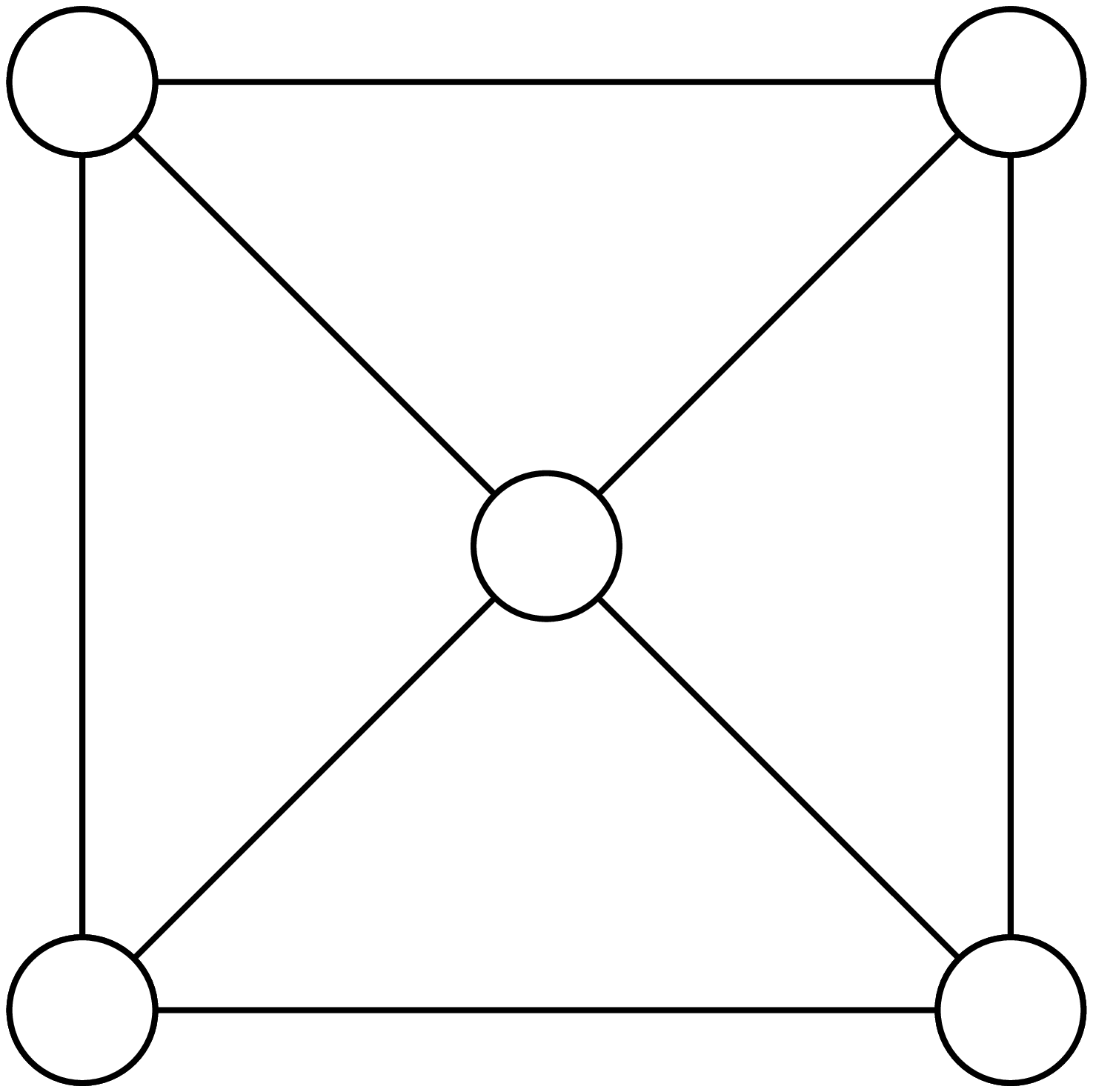}\\
{\rm (a)} &{\rm (b)} 
\end{array}
$$
\caption{Lattices: (a) five sites and six bonds; 
(b) lattice (a) with two additional bonds.}      
\label{fig:lattice5}
\end{figure}

\begin{figure}[t]
$$\begin{array}{cc}
{\rm (a)} & \includegraphics[width=.50\textwidth]{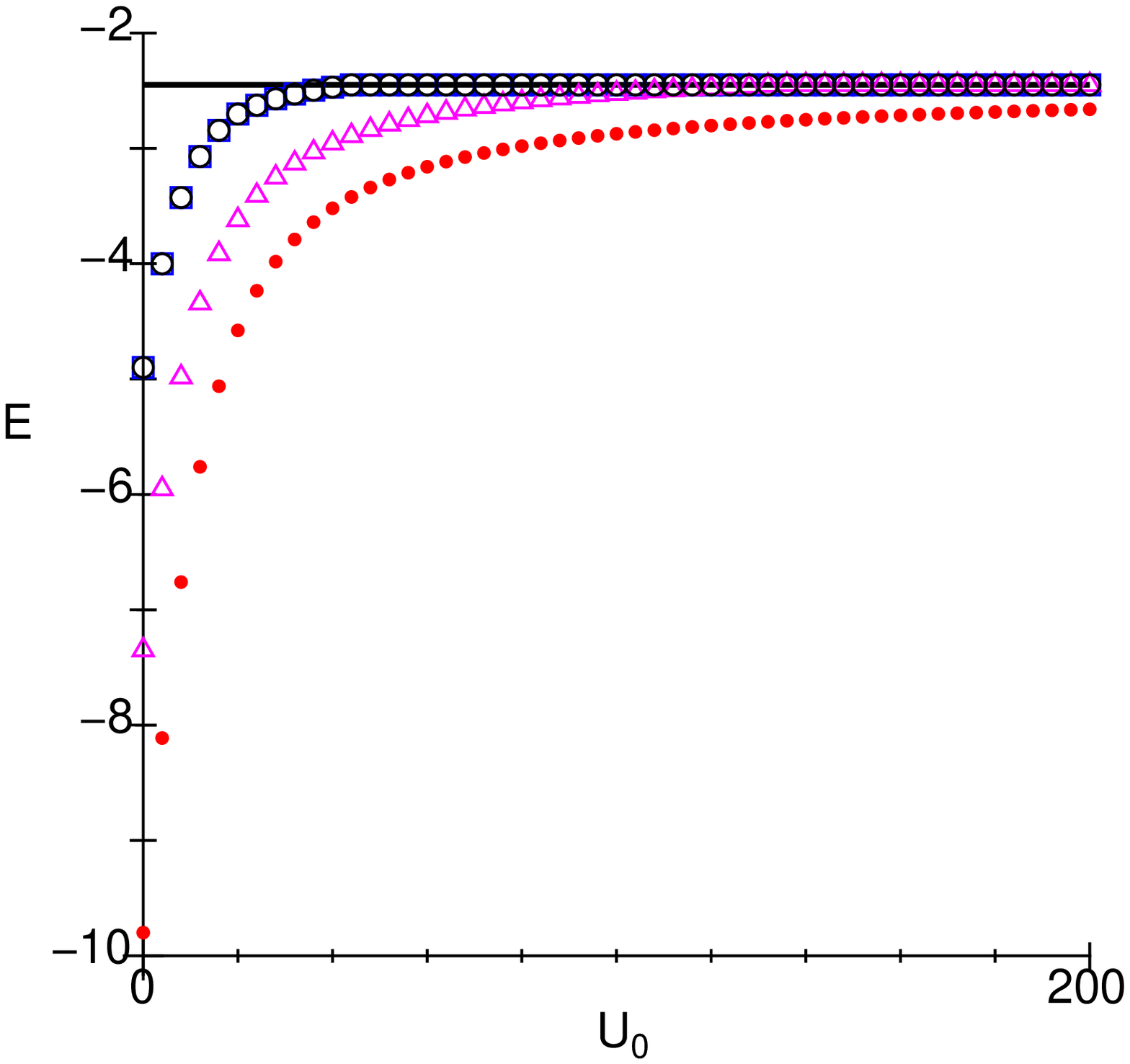}\\
{\rm (b)} & \includegraphics[width=.50\textwidth]{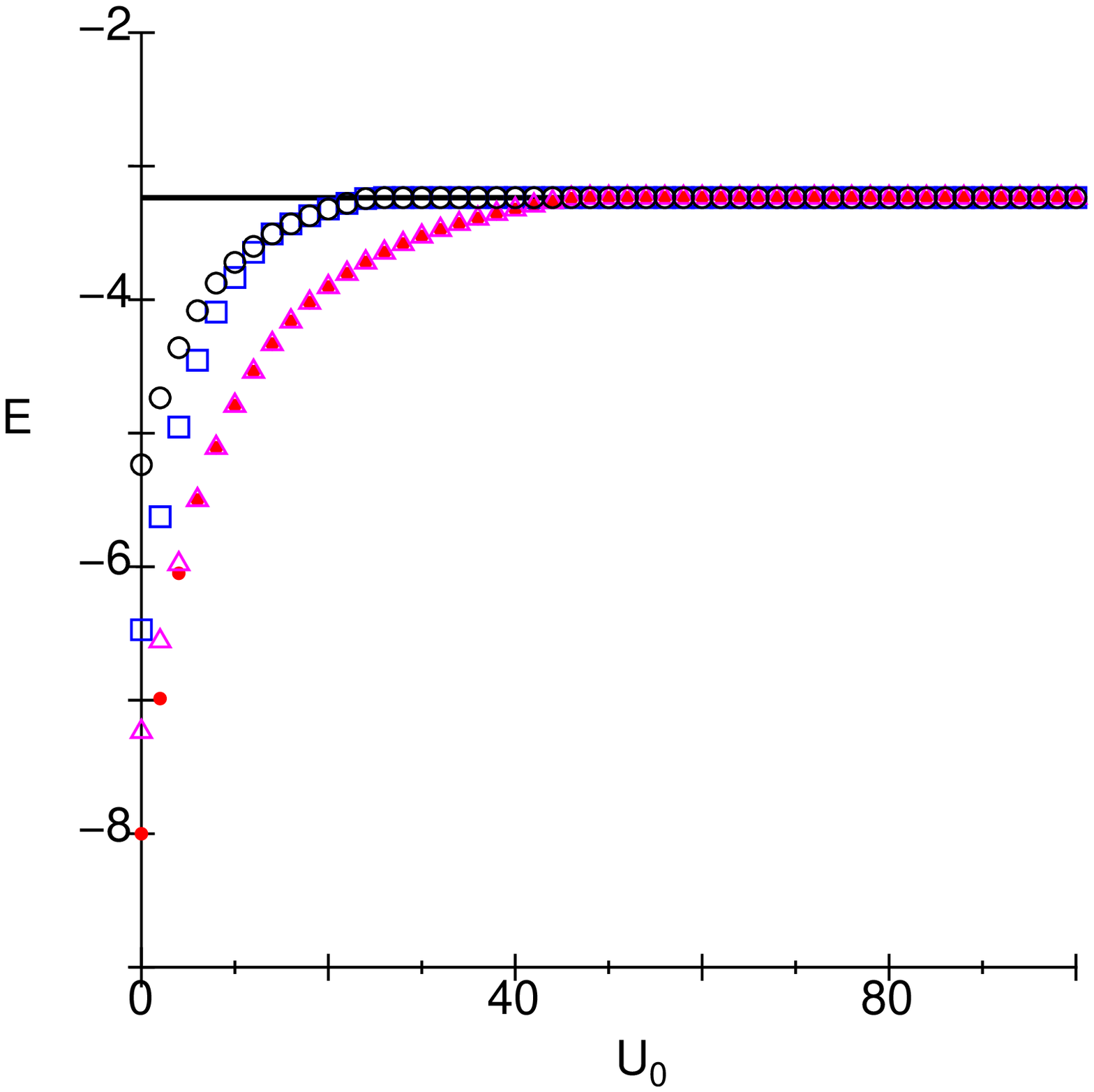}
\end{array}
$$
\caption{Energies of the ground state 
for 
several sets of $(n_{3/2},n_{1/2},n_{-1/2},n_{-3/2})$.
(a) A system with four 
particles on the lattice shown in Fig.~\ref{fig:lattice5}(a). 
The solid line denotes the ground state energy for $(4,0,0,0)$, 
circles: (3,1,0,0),
squares: (2,2,0,0),
triangles: (2,1,1,0), and
bullets: (1,1,1,1).
(b)  Same as (a) but for the lattice shown in Fig.~\ref{fig:lattice5}(b).  
}
\label{fig:energy3}
\end{figure}

Let us now study the dependence of the ground state energy on $U_0$.
We consider a system with four particles with $S=3/2$
on the 5-site lattice depicted in Fig.~\ref{fig:lattice5}(a).
Figure~\ref{fig:energy3}(a)
shows the ground state energies for the sets 
$\{n_M\}$=$(n_{3/2},n_{1/2},n_{-1/2},n_{-3/2})$ =
$(4,0,0,0)$, $(3,1,0,0)$, $(2,2,0,0)$,
$(2,1,1,0)$, $(1,1,1,1)$,
corresponding to systems in which the number of different states
is 5, 50, 100, 250 and 625, respectively.
The Hamiltonian does not depend on the values of $M$, and thus the energy 
level structure is same for the cases with the same set of numbers,
e.g. for (3,1,0,0),(3,0,1,0),$\cdots$, (0,0,1,3).
For small value of $U_0$, the ground state energies are all different.
As $U_0$ increases, the ground state energies of (3,1,0,0), (2,2,0,0) and (2,1,1,0)
become degenerate with that of (4,0,0,0). 
This fact indicates that the lowest energy state in these sets
has the total spin $S_{\rm max}$.
However, we find that the lowest energy state in the set (1,1,1,1) 
is always lower than that of (4,0,0,0). 
This means that even at large $U_0$ the ground state has total spin $S<S_{\rm max}$.
Therefore, the Nagaoka-ferromagnetic state, i.e., the state of the maximum total
spin, is not realized 
as the ground state in the present case even at large values of $U_0$.

In general, when there are multiple conserved quantities in a system 
that do not commute with each other, each energy state of the system
is usually degenerate as shown in Eq.~(\ref{HEG})~\cite{Landau}.
In the present model, the total magnetization and the set $\{n_M\}$ are conserved. 
However, the ground state in Fig.~2(a) is not degenerate, which gives an exception to
the above general property. 

Figure 3 shows the $U_0$ dependence of the total spin 
of the lowest-energy state for the sets $\{n_M\}$ 
on the lattice of 
Fig.~\ref{fig:energy3}(a).
The total spin of states in the subspace (4,0,0,0) is 
$S_{\rm max}=(3/2)\times 4=6$. For small values of $U$,
the total spin of states in the subspace (3,1,0,0) is less than  $S_{\rm max}$, but
it becomes $S_{\rm max}$ when the ground state energy becomes degenerate to that of 
the (4,0,0,0) subspace.
The ground state energies of (2,2,0,0) and (2,1,1,0) 
become degenerate to that of (4,0,0,0) 
at certain values of $U_0$. 
However, the total spin of the lowest-energy state of these sets does not 
reach $S_{\rm max}$. 
This fact agrees with the earlier argument that 
if $S^-|G,M\rangle$ consists of more than one set of $\{n_M\}$'s,
neither of these sets yields an eigenstate of the total spin although
each of them have the same ground state energy.
Therefore, the expectation value of the total spin is not necessarily an integer. 

\begin{figure}[t]
$$\begin{array}{c}
\includegraphics[width=.5\textwidth]{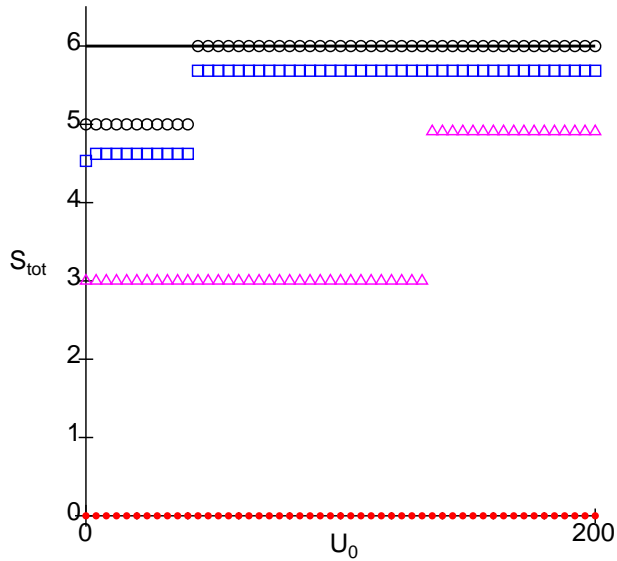}
\end{array}
$$
\caption{The total spin $S_{\rm tot}$ of the lowest-energy state 
for the cases shown in Fig.\ref{fig:energy3}(a).
The legend is the same as in Fig.2.  
}
\label{fig:totalS}
\end{figure}

\begin{figure}[t]
$$\begin{array}{cc}
\includegraphics[width=.20\textwidth]{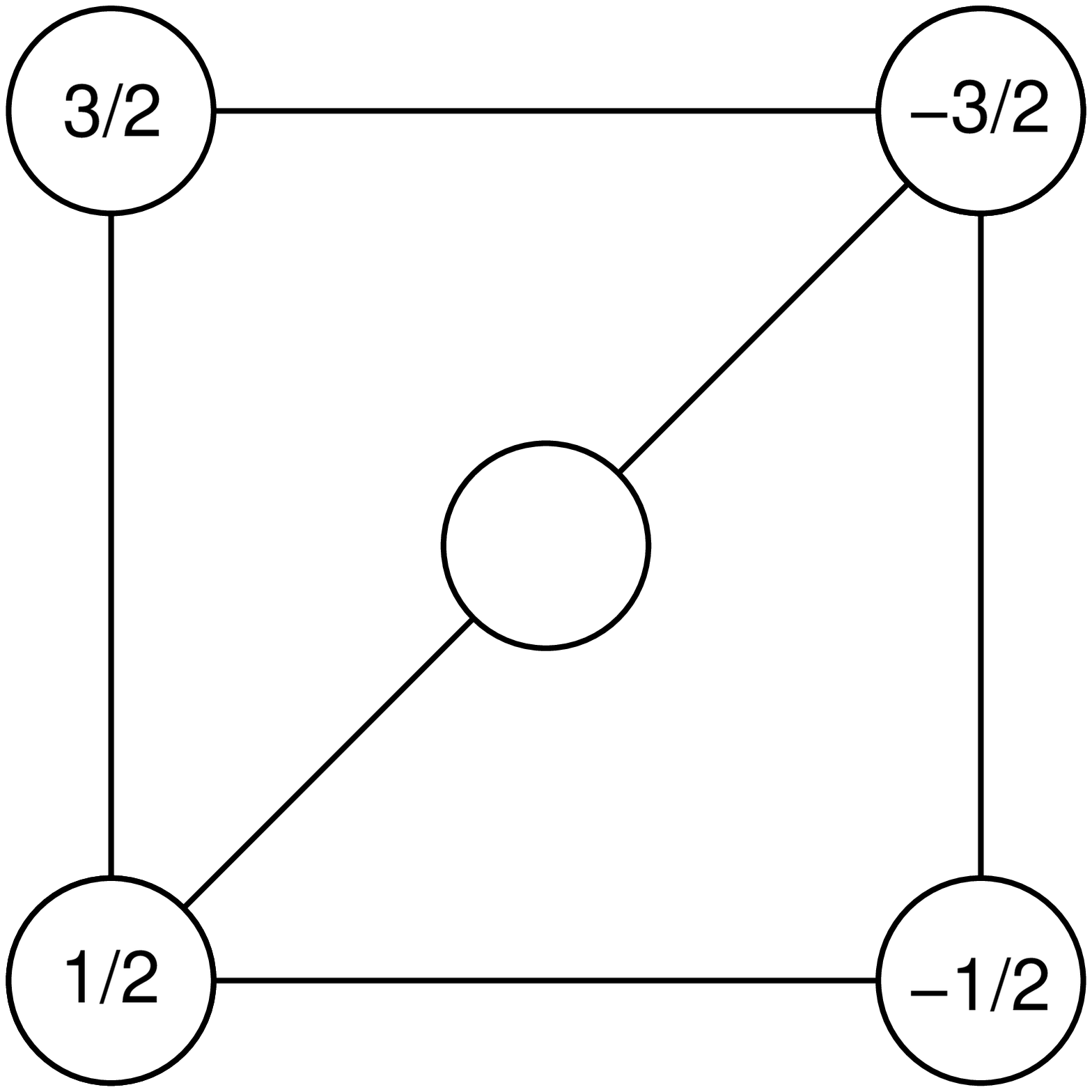}
&
\includegraphics[width=.20\textwidth]{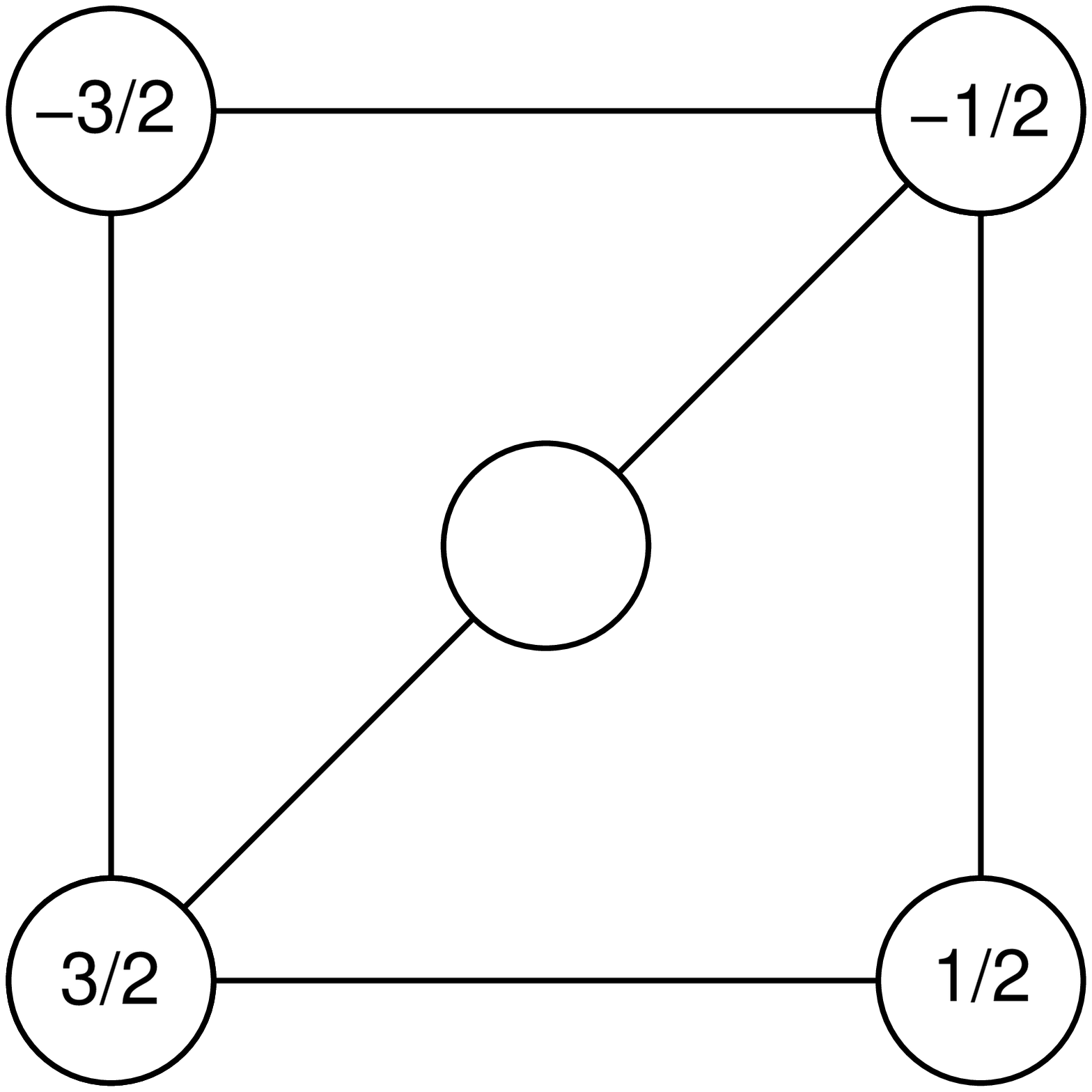}\\
{\rm (a)} &{\rm (b)}\end{array}
$$
\caption{Nearest-neighbor hopping cannot change 
configuration (a) into (b) without creating double occupancy.}
\label{fig:conf}
\end{figure}

\subsection{Connectivity condition for the system with $S=\frac{3}{2}$}

Note that the lattice shown in Fig.~\ref{fig:lattice5}(a) satisfies 
the so-called connectivity condition for the case of $S=1/2$
and the Nagaoka-ferromagnetic state is realized at large $U_0$.
However, for $S=3/2$, this state is not realized.
The reason is that the connectivity condition for 
four $S=3/2$ particles with different magnetization, i.e., $M=-3/2,-1/2,1/2,3/2$
is not satisfied
in the lattice of Fig.~\ref{fig:lattice5}(a).
Indeed, the configuration depicted in Fig.~\ref{fig:conf}(a) can not be changed into
that of Fig.~\ref{fig:conf}(b).

In order to re-establish the connectivity condition we may add bonds and construct 
the lattice depicted in Fig.~\ref{fig:lattice5}(b). 
This lattice has loops with an odd number of 
bonds and therefore we need to use a negative value of $t$. Then, we find
that the Nagaoka-ferromagnetic state is realized, as is clear 
from Fig.~\ref{fig:energy3}(b).
Summarizing, we have shown that $S=3/2$ itinerant particle systems 
may exhibit Nagaoka-ferromagnetism although 
it is more difficult to satisfy the connectivity condition.

\subsection{Structure of the ground state}

We found that the ground state in each subspace for a set $\{n_M\}$
is unique but in each subspace we have one state with the same ground-state energy.
In Table~\ref{table1}, we list all the sets $\{n_M\}$.
The structure indicates that there is a state with $S_{\rm tot}=6$,4, 2 and 0.
In the case of $S=1/2$, 
there is a one-to-one correspondence between the magnetization $M$ and $\{n_+,n_-\}$, 
as is clear from Table~\ref{table1}.
Thus, when we create states by applying $S^-$ from the all-up state,
we have only the state with $S_{\rm tot}=2$ 
for the four spin system.
In contrast, in the case of $S=3/2$, we create different sets by applying
$S^-$. As we mentioned in the previous section, the created states are degenerate
as ground states. Therefore, the structure displayed in Table~\ref{table1} is
intrinsic for systems with $S=3/2,5/2,\ldots$.

\begin{table}
\caption{Sets of particles of different $M(\ge 0)$ $\{n_M\}$ 
for the cases of $S=3/2$ and $1/2$.
For $S=3/2$, we list the sets only for positive $M$.}
\label{table1}
\begin{tabular}{c|ccccccc}
$S=3/2$ &&&&&&&\\
\hline
$M$      & 6       & 5       & 4       & 3       & 2       & 1         & 0 \\
\hline
$\{n_M\}$&(4,0,0,0)&(3,1,0,0)&(3,0,1,0)&(3,0,0,1)&(0,4,0,0)& (0,3,1,0) & (1,0,3,0) \\
       &          &         &(2,2,0,0)&(1,3,0,0)&(2,1,0,1)& (2,0,1,1) & (0,3,0,1) \\
       &          &         &         &(2,1,1,0)&(1,2,1,0)& (1,2,0,1) & (2,0,0,2) \\
       &          &         &         &         &(2,0,2,0)& (1,1,2,0) & (0,2,2,0) \\
       &          &         &         &         &         &           & (1,1,1,1) \\
\hline
&&&&&&&\\
$S=1/2$ &&&&&&&\\
\hline
$M$      & 2       & 1       & 0       & -1       & 2       &          &  \\
\hline
$\{n_M\}$&(4,0)&(3,1)&(2,2)&(1,3)&(0,4)& &\\
\hline
\end{tabular}
\end{table}

\section{Ground state of a boson system with $S=1$}

\begin{figure}[t]
\includegraphics[width=.5\textwidth]{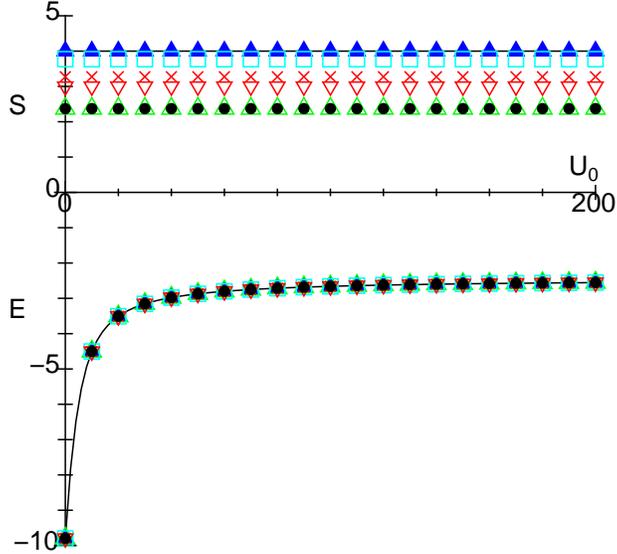}
\caption{Ground state energies $E$ and the corresponding total spin $S_{\rm tot}$ 
as a function of $U_0$ 
for four $S=1$ bosons and various $(n_1,n_0,n_{-1})$ on the lattice shown 
in Fig.~\ref{fig:lattice5}(a).
The solid line denotes data for (4,0,0), 
solid triangles: (3,1,0),
open triangles: (3,0,1), 
open squares: (2,2,0),
reversed triangles: (2,1,1),
crosses: (1,2,1), and
bullets: (0,4,0).
}
\label{fig:energyB}
\end{figure}

Next, we study a system of $S=1$ particles, namely a boson system.
Figure~\ref{fig:energyB}
shows the ground state energies for the sets $\{n_M\}=(n_1,n_0,n_{-1})$=
$(4,0,0)$, $(0,4,0)$, 
$(3,1,0)$, $(3,0,1)$,
$(2,2,0)$, 
$(2,1,1)$,
$(1,2,1)$,
corresponding to systems in which the number of different states
is 70, 70, 175, 175, 225, 375, and 375, respectively.
Surprisingly, 
we find that the energies are the same for all the cases regardless of $U_0$.
The total spin of the different ground states are also plotted in Fig.~\ref{fig:energyB}.
Thus, the magnetic properties of the boson system are very different 
from that of the fermion system.
The total spins are less 
than the maximum value ($S_{\rm tot}$=4) except for (4,0,0) and (3,1,0)
which reflects the fact that the eigenstate of maximum total spin 
contains more than two sets of $\{n_M \}$, as in the case of the $S=3/2$ system. 

We find a ground state in each set as in the case of $S=3/2$.
We list the sets in Table~\ref{table2}.
\begin{table}
\caption{Sets of particles of different $M(\ge 0)$ $\{n_M\}$ for the case of $S=1$.
}
\label{table2}
\begin{tabular}{c|ccccccc}
$S=1$ &&&&&&&\\
\hline
$M$      & 4       & 3       & 2       & 1         & 0 \\
\hline
$\{n_M\}$&(4,0,0)&(3,1,0)&(3,0,1)&(2,1,1)&(0,4,0)\\
       &          &     &(2,2,0)&(1,3,0)&(2,0,2)\\
       &          &         &     &      &(1,2,1)\\
\hline
\end{tabular}
\end{table}
Here we find again a degenerate ground states with 
$S_{\rm tot}=4,2$ and 0, 
which is an intrinsic property of systems with $S=1,2,\ldots$.

Moreover, we find that even in the half-filled case, 
in the boson system the total spin takes the maximum value. 
For instance, for a system of four atoms on a
simple square lattice (corresponding to the half-filled case), the 
ground state is the same for all sets $\{ n_M \}$ (results not shown).

This property of boson systems can be understood as follows.
Consider the subspace for a fixed set $\{n_M\}$. 
All the off-diagonal matrix elements of the Hamiltonian ${\cal H}$ are $-t$ or 0. 
By substracting an appropriate multiple of the unit matrix, 
also the diagonal elements can be made negative or zero.
Thus, all the elements of the shifted matrix ${\widetilde H}$ are either negative or zero.
In our model, all the sites are connected by bonds and therefore
there exist a number $n>0$ such that all the elements of ${\widetilde H}^{2n}$ are positive.
Then, the Perron-Frobenius theorem tells us that there exist
a unique eigenstate of ${\widetilde H}^{2n}$ with an eigenvalue that is larger than
the absolute value of all other eigenvalues.
This unique eigenstate is therefore the ground state of $H$,
is totally symmetric with positive coefficients, and contains the state 
of the maximum total spin. 
However, as we mentioned above, this ground state has an intrinsic degeneracy with respect
to subspaces that have different $\{n_M\}$ because we always start from the state 
with all spins maximum and let $S^-$ create different sets of $\{n^\prime_M\}$.
The states that are generated in this manner have all positive coefficients
and are therefore ground states too.
Cleary, this property does not depend on the connectivity of the lattice 
or the value of $U_0$.

In contrast, for fermion systems, it is in general impossible to
transform $H$ such that all elements have the same sign but
in those cases for which such a transformation exist, which is precisely                  
the condition of Nagaoka ferromagnetism, we can apply the same arguments as in the boson
case to prove that the ground state of $H$
is totally symmetric with positive coefficients and contains the state of 
the maximum total spin. 

\section{Summary and Discussion}

We have studied the magnetic properties of the ground state of itinerant systems with $S>1$. 
We found that fermion systems ($S=3/2$) support an extended form
of Nagaoka ferromagnetism but that the connectivity condition is more
difficult to satisfy because of the presence of particles with different magnetization.
When the maximum $S_{\rm tot}$ state is the ground state, the system has a 
degenerate ground state in each set $\{n_M\}$ listed in Table~\ref{table1}.
Thus the ground state manifold consists of not only the state of the maximum $S_{\rm tot}$ 
but also of states with smaller values of $S_{\rm tot}$. 
This degenerate structure is intrinsic
for the systems with $S>1/2$ where several sets of $\{n_M\}$ exist for a given value of $M$.

For boson systems ($S=1$) we found that there is a ground state in each set of $\{n_M\}$ 
and that the same type of degenerate ground-state structure appears 
as the one found for $S=3/2$, but in this case
regardless of the value of $U_0$ and of the shape of the lattice, which is 
an intrinsic property of the bosonic case. 
Thus, we conclude that boson itinerant magnetic systems always 
have a state with the maximum total spin belonging to the manifold of ground states.
This property follows quite naturally from the fact that  
the Hamiltonian of the boson system can be transformed such that 
all elements have the same sign, implying that the ground state is fully symmetrized.
In contrast, for fermion systems we need an additional condition, the condition for the
Nagaoka ferromagetism, for the ground state to have maximum total spin.

We expect that these properties will be confirmed in real experimental 
systems.
 
\section*{Acknowledgement}

The present work was supported by Grant-in-Aid for Scientific Research
on Priority Areas, and also and the Next Generation Super Computer
Project, Nanoscience Program from MEXT of Japan.
The numerical calculations were supported by the supercomputer center of
ISSP of Tokyo University.
Partial support by the NCF, The Netherlands is gratefully acknowledged.

\end{document}